\documentstyle[twoside,fleqn,espcrc2]{article}%
\newcommand{\beq}{\begin{equation}}
\newcommand{\eeq}{\end{equation}}
\newcommand{\bea}{\begin{eqnarray}}
\newcommand{\eea}{\end{eqnarray}}
\begin{document}
\renewcommand{\refname}{\normalsize\bf References}
\title{
Exotic Stochastic Processes from Complex Quantum Environments}

\author{%
   Dimitri KUSNEZOV%
        \address{Center for Theoretical Physics,
                 Sloane Physics Laboratory\\
                 Yale University, New Haven, Connecticut 06520-8120,
                 USA}
\thanks{The present work has been partially supported by DOE.
        The Laboratoire de Physique Th\'eorique is a Unit\'e Mixte
        de Recherche du C.N.R.S., UMR 8627},
         Aurel BULGAC%
        \address{Department of Physics,
                 University of Washington\\
                 P.O. Box 351560, Seattle, Washington 98195--1560, USA}%
\thanks{Address for the period October 1, 1999 -- June 30, 2000,
        Max--Planck--Institut f\"ur Kernphysik, Postfach 10 39 80,
        69029 Heidelberg, GERMANY. },
        \,and
        \,Giu DO DANG%
        \address{Laboratoire de Physique Th\'eorique,
                 Universit\'e de Paris--Sud\\
                 B\^atiment 211, 91405, Orsay, FRANCE},%
     }
%
%
\begin{abstract}
\hrule
\mbox{}\\[-0.2cm]

\noindent{\bf Abstract}\\
Stochastic processes are shown to emerge from the time evolution of
complex quantum systems. Using parametric, banded random matrix
ensembles to describe a quantum chaotic environment, we show that the
dynamical evolution of a particle coupled to such environments
displays a variety of stochastic behaviors, ranging from turbulent
diffusion to L\'evy processes and Brownian motion.  Dissipation and
diffusion emerge naturally in the stochastic interpretation of the
dynamics. This approach provides a derivation of a fractional kinetic
theory in the classical limit and leads to classical L\'evy dynamics.
\\[0.2cm]

\hrule
\end{abstract}

\maketitle

\section{Introduction}

The understanding of how stochastic processes emerge from classical
dynamical systems is closely related to classical chaos. Often one
finds that the dynamics is non--Gaussian, displaying either enhanced or
dispersive behavior \cite{szk}. One can find such behavior in the
interaction of slow and fast degrees of freedom in many--body
systems \cite{us}, in tracer diffusion in turbulent backgrounds such as
the atmosphere, or random potentials, and many more \cite{kzs}.  The
wide variety of processes which exhibit anomalous behavior in the
transport has led to a variety of theoretical efforts, including
fractional extensions of kinetic theory \cite{zas,fp,klafter}, random
walks in random potentials \cite{bouch}, power law noise in generalized
Langevin equations \cite{fog}, stochastic webs \cite{zas} and L\'evy
walks and flights \cite{mon,kzs}. While the common thread to these
approaches are generalizations of Brownian motion known as L\'evy
stable laws (discussed below), there is no common theoretical
foundation.  For instance, fractional kinetic equations are postulated
in such a manner as to provide the desired diffusion through scaling
arguments.

In turning to the quantum theory, we might ask whether it is possible
to realize anomalous diffusion or L\'evy stable laws in the time
evolution of quantum Hamiltonians \cite{prl}. If classical chaos is the
origin of the stochastic processes in classical dynamical systems, it
is natural to ask whether quantum chaos can be the source of quantum
stochastic processes. Starting with the quantum counterpart of
classical chaos, namely random matrix theory (RMT), we consider
whether the time evolution of such Hamiltonians can generate
stochastic processes. We will see that not only is this possible, but
depending on certain properties of the quantum Hamiltonian, one can
realize a full range of stochastic processes, including those of
L\'evy.  Further, fractional kinetic theory develops naturally in the
semi--classical limit.

To understand diffusion observed in many systems which have $\langle
R^2(t)\rangle \sim t^\gamma$, where $\gamma\not= 1$, generalizations
of Brownian motion have been sought. L\'evy processes provide such an
approach. In studies of the extensions of the central limit theorem,
P.  L\'evy found a continuous class of non--Gaussian processes that
satisfy the same fundamental equation that gives rise to the theory of
Gaussian processes \cite{levy}. In one dimension, the L\'evy stable
laws have the form
\begin{equation}
P(x,t)={\cal L}_\alpha ^A(x)=\int\frac {dk}{2\pi } \exp\left\{ikx-A|k|^\alpha
\right\}
\end{equation}
where $0<\alpha \leq 2$ and $A\propto t$.  Gaussian processes
correspond to the case where $\alpha =2$.  The L\'evy distributions
are scale invariant,
\begin{equation}
{\cal L}_\alpha ^A(x)=A^{-1/\alpha }{\cal L}_\alpha ^1(xA^{-1/\alpha }),
\end{equation}
where for $A=1$ we drop the superscript: ${\cal L}_\alpha ^1(x)={\cal
L}_\alpha (x)$. The scale invariance indicates that the trajectory
followed by the random process will not be dominated by one
characteristic scale, resulting in a self--similar behavior.  With the
exception of the Gaussian ($\alpha=2$), all the L\'evy stable laws
have infinite second moments \cite{mon}.  For certain values of
$\alpha$, it is possible to compute the inverse Fourier transform in
(1) \cite{prague}.

\section{Quantum Chaotic Environments} 

We would like to understand the diffusion of a quantum particle
interacting with a quantum chaotic environment. Random matrix theory
provides a convenient description of quantum chaotic systems, but as
it contains no scales or physics, we must develop the notion of a
random environment. Since our approach is Hamiltonian based, we
envision the particle interacting with a complex quantum system, which
might have a non--trivial density of states, and which is inherently
chaotic in the sense of RMT fluctuations of the matrix elements. We
will assume that the Hamiltonian $H_e$ which describes this
`environment' is time--reversal invariant, so that it is a real,
symmetric matrix. (This serves only to simplify the notation). As the
background is chaotic, it is not necessarily thermal.

The Hamiltonian for the chaotic environment plus interaction will have
the form
\begin{equation}
H_e=h_0(x,p)+h_1(X,x,p).
\end{equation}
where $(x,p)$ are the coordinates and momenta of the environment, and
$(X,P)$ are those for the test particle.  $h_0$ is the Hamiltonian of
the environment and $h_1$ is the interaction with the test
particle. It is convenient to choose the fixed basis of $h_0$ to
describe the matrix elements of $h_1$. In this basis, we denote
\begin{equation}
h_0\mid n\rangle=\varepsilon _n\mid  n\rangle ,\qquad (n=1,...,N),
\end{equation}
so that the matrix elements of $H_e$ are
\begin{equation}
\lbrack H_e]_{ij}=\varepsilon _i\delta _{ij}+[h_1(X)]_{ij}.
\end{equation}
There is now control over the density of states of $h_0$ by suitably
choosing the $\varepsilon_i$ to describe the system of interest. We
would like to introduce a parameter related to the average level
density ($\rho(\varepsilon)$) of $h_0$ as $\beta=1/T=d\ell n
\rho(\varepsilon)/d\varepsilon$, or equivalently,
$\rho(\varepsilon)=\rho_0\exp(\beta \varepsilon)$. We will see that
this derivative of the level density naturally plays the role of
temperature in the quantum dynamics in certain cases.  We now choose
$h_1$ to be chaotic, which allow us to apply the Gaussian orthogonal
ensemble (GOE) to the matrix elements. In this case, the chaotic
properties of the interaction between the environment and the test
particle are built into the correlation function (second cumulant):
\begin{equation}\label{eq:secmom}
\langle [h_1(X)]_{ij}[h_1(Y)]_{kl}\rangle
={\cal G}_{ij}(X-Y)\Delta_{ijkl}.
\end{equation}
Here $\Delta_{ijkl}=[\delta_{ik}\delta _{jl}+\delta _{il}\delta
_{jk}]$, and all other cumulants vanish.  A convenient realization of
the correlation function is \cite{brink,bdk1,bdk2}:
\begin{eqnarray}
{\cal G}_{ij}(X)&=&
\frac{\Gamma ^\downarrow }{2\pi \sqrt{\rho (\varepsilon _i)\rho (\varepsilon
_j)}}\nonumber \\
&&\times 
\exp \left [ -\frac{(\varepsilon _i -\varepsilon _j)^2}{2\kappa _0
^2} \right ]G \left (\frac{X}{X_0}\right ).
\end{eqnarray}
The essential elements of this function are the following. We expect
due to selection rules that $h_1$ is not a full matrix, but most
likely banded in some way. Here the band width is described through
$\kappa _0$. As the energy increases, the level density typically
increases rapidly, so one expects the average interaction matrix
elements to reduce, which is accounted for by the $\rho^{-1/2}$
factors.  One also expects $H_e$ to decorrelate in $X$ on some length
scale $X_0$ (NB This is not generally the scale on which the
eigenvalues of $H_e(X)$ vary as suggested by the avoided level
crossings).  Finally, $G(x)=G(-x)=G^*(x)\le 1$, $G(0)=1$, and the
overall strength $\Gamma ^\downarrow$ is known as the spreading
width. Before we turn to the dynamics, let us consider the statistical
nature of $h_1$.

\section{Short Distance Correlations} 

An interesting consequence of the statistical measure leading to (6)
is that the correlation function given by the two--point function must
decorrelate no faster than quadratic \cite{caio}.
In many cases, the essential characteristic of $G(x)$ is governed by
its short distance behavior.  On short distance scales we approximate:
\begin{equation}
G(x)\approx 1-c_\alpha \left| x\right| ^\alpha + \ldots
\end{equation}
For systems with smooth parameter dependence, one has $\alpha=2$. A
Dyson process on the other hand has $\alpha=1$ \cite{caio}. Similarly,
disordered systems are often modelled with $\alpha=1$, for instance
when one uses $\langle V(x) V(x')\rangle\sim \exp[-|x-x'|]$.

From the requirement that the probability distribution of matrix
elements $[h_1(X)]_{ij}$ be always a positive definite function and 
using the theorems of Bochner, it can be shown that the
correlator $G(x)$ has to satisfy the following restriction at short distances
\begin{equation}
  0< \alpha \le 2.
\end{equation}
Hence, we obtain a finite range of values $\alpha$ which are allowed
for parametric random matrix ensembles.

As the position $X$ of the slow particle changes, the instantaneous
energy levels $E_n(X)$ of $[h_1(X)]_{ij}$ change.  Using the above
expresion for the correlator $G(x)$ one obtains that
the average fluctuations are
\begin{equation}
\langle\left [ E_n(X) -E_n(Y) \right ] ^2
 \rangle = D_\alpha |X-Y|^\alpha.
\end{equation}
The energy--spacing fluctuations have a behavior, which is similar to
a L\'evy process characterized by the diffusion constant
$D_\alpha$. The character of these fluctuations in the eigenvalues
$E_n(X)$, indicated by $\alpha$, will be seen to be related to L\'evy
distributions, which describe the time evolution of the density matrix
for a particle evolving in this chaotic bath.

\section{Quantum L\'evy Processes} 

We now derive the dynamics of a quantum particle coupled to this
quantum chaotic environment. We use the Hamiltonian
\begin{equation}
H_{ij}(X,P)=\delta _{ij}\left [ \frac{P^2}{2M}+U(X)\right ] +H_{e,ij}(X).
\end{equation}
We will be interested in the diffusive and dissipative behavior
of the test particle induced by the environment, so we will
neglect the effects of the potential $U(X)$.  The evolution
equation for the density matrix of the test particle for this
class of RMT Hamiltonians has been derived recently using
influence functional techniques \cite{bdk1}. It is given in a
high temperature expansion, and up to $o(\beta)$ has the form:

$$
i\hbar \frac{\partial\rho (X,Y,t)}{\partial t}
  =\left\{ \frac{P_X^2}{2M}-\frac{P_Y^2}{2M}
+U(X)-U(Y)\right.
$$
\begin{eqnarray}
&&-\frac{\beta \Gamma ^{\downarrow }\hbar }{4X_0M}G^{\prime }\left( \frac{%
X-Y }{X_0}\right) (P_X-P_Y)  \nonumber \\
&&\left. +i\Gamma ^{\downarrow }\left[ G\left( \frac{X-Y}{X_0}\right)
-1\right] \right\} \rho (X,Y,t).
\end{eqnarray}
To extract the dynamics analytically, we pass to the weak--coupling
limit \cite{bdk1}, in which we keep only the leading order term in the
correlation function. This corresponds physically to the case where
the test particle does not provide significant feedback to the
environment.  In this limit, we have $G(x)=1-|x|^\alpha$ so that
$G'(x)$ represents $-\alpha{\rm sign}(x) |x|^{\alpha-1}$.

In the absence of a potential $U(X)$, and when the level density is
constant in the region of interest ($\beta=0$), these equations are
readily solved for the test particle density matrix $\rho(X,Y,t)$. In
the variables $r=(X+X^{\prime })/2$, $s=X-X^{\prime }$, the solution
is
$$\rho (r,s,t) =\int dr^{\prime }\int \frac{dk}{2\pi \hbar }\rho
_0\left ( r^{\prime },s-\frac{kt}{m}\right )
$$
\begin{eqnarray}
&&\times\exp \left[ \frac{ik(r-r^{\prime })}\hbar \right.
\nonumber \\
&&\left. +\frac{\Gamma ^{\downarrow }M}{\hbar k}
\int_{s-kt/M}^sds^{\prime}
\left [ G\left ( \frac{s^{\prime }}X_0 \right ) -1\right ]\right]  \label{rho}
\end{eqnarray}
The initial density matrix at $t=0$ is denoted by
$\rho(X,Y,0)=\rho_0(X,Y)$.

The probability of finding the test particle at a position $X$ at a
certain time $t$, denoted $P(X,t)$, is given by the diagonal elements
of the density matrix: $P(X,t)=\rho(X,X,t)=\rho (r,s=0,t)$.  This is
readily computed from Eq. (20) as: 
$$
\rho (r,0,t) =\int 
\frac{dk}{2\pi \hbar }\exp \left[ -k^2\left[ \frac{\sigma ^2}{2\hbar
^2}+\frac{t^2}{8M\sigma ^2}\right] \right. 
$$
\begin{equation}
\left. -\frac{\Gamma ^{\downarrow }t^{\alpha +1}}{(\alpha +1)\hbar
(MX_0)^\alpha }|k|^\alpha +ik\frac r\hbar \right] .  \label{eq:levy}
\end{equation}
We have assumed that the test particle is initially in a Gaussian
wave-packet
\begin{eqnarray}
\psi(X)&=&\exp[-X^2/4\sigma^2]/[2\pi\sigma^2]^{1/4}\\
\rho_0(r,s)&=&\frac{1}{\sqrt{2\pi}\sigma}\exp\left[-\frac{1}{8\sigma^2}
  (4r^2+s^2)\right].
\end{eqnarray}

Comparing to Eq. (1), we see that Eq. (21) is a Fourier
transform of a Gaussian and a L\'evy process, which can be
expressed as the convolution of the two distributions,
\begin{equation}\label{eq:px}
P (X,t)=\int dX^{\prime }{\cal L}_\alpha ^{a(t)}(X^{\prime })
{\cal L}_2^{b(t)}(X-X^{\prime }),
\end{equation}
with
\begin{eqnarray}
 a(t) &=&\frac{\Gamma ^{\downarrow }}{(\alpha +1)\hbar }\left( \frac \hbar {
               MX_0}\right) ^\alpha t^{\alpha +1}, \\
 b(t) &=&\frac{\sigma ^2}2+\frac{\hbar ^2}{8M^2\sigma ^2}t^2.
\end{eqnarray}
It is interesting to note that these processes are Markovian, having
no memory effects which are responsible for the anomalous character.
The dynamics of these processes are now labeled by $\alpha$, which was
shown in Section 3 to be in the range $0< \alpha \leq 2$. Hence,
Eq. (17) shows us that the short distance decorrelations of the
adiabatic states of $H_e$ are directly related to the type of
stochastic process which develops in the quantum time evolution of the
test particle.

\section{Turbulent--like Diffusion }

For the cases $\alpha<1$ and $\alpha >1$, the character of these
L\'evy processes has been discussed in \cite{prl}. When $\alpha=2$ and
$\beta=0$, $P(X,t)$ is the convolution of two Gaussians, and as a
consequence is Gaussian as well. The test particle in this case
exhibits turbulent diffusion, with
\begin{eqnarray}
\left\langle  X^2;t\right\rangle  &=&\int
dX\;X^2\rho(X,X,t)\\
\  &=&\sigma ^2+\frac{\hbar ^2}{4M^2\sigma ^2}t^2+\frac{\Gamma ^{\downarrow
}\hbar }{3M^2X_0^2}t^3  \label{rr}
\end{eqnarray}
Here one can see that the initial width at $t=0$ is due to the
Gaussian initial condition, and the second term reflects the natural
spreading of a wave--packet in the absence of an environment. The $t^3$
character of the dissipative contribution which arises from the
environment is typical of turbulent backgrounds. The range of times
under which we expect such diffusion to hold has been discussed in
Ref. \cite{pla}.  In Table 1 we summarize selected properties of the
RMT influence functional and its stochastic behaviors.

\section{Fractional Kinetic Theory} 

One of the phenomenological approaches to anomalous diffusion is to
postulate a Fokker--Planck (FP) equation which supports the types

of diffusion seen in complex classical systems. To this end, a number
of extensions of the FP equation have been proposed in which the
spatial and/or time derivatives are replaced with derivatives of
fractional order. Fractional derivative/integrals are typically
defined through the use of integral transforms \cite{samko}. In one of
the many realizations, for example, one can define ${\cal D}^\alpha_x$
to be a derivative (integral) of order $\alpha$ when $\alpha>0$
($\alpha<0$) as follows. For $\alpha>0$ let
\begin{equation}
({\cal D}_t^{-\alpha }f)(x)={\frac 1{\Gamma (\alpha )}}\int_0^xdy(x-y)^{
\alpha -1}f(y),
\end{equation}
where $\alpha >0$ is real, $f(x)$ is an arbitrary function and $\Gamma
(\alpha )$ is the $\Gamma $--function. For $f(x)=x^\mu $, where $\mu
>0$ is real, this gives
\begin{equation}
{\cal D}_x^{-\alpha }x^\mu =\frac{\Gamma (\mu +1)}{\Gamma (\mu +1+\alpha )}
t^{\mu +\alpha }.
\end{equation}
Clearly for $\alpha =n$ a positive integer, this is just the $n$--th
iterated integral of $f(x)$. For general $\alpha>0$ it is a suitable
definition for the fractional integral of order $\alpha$. For negative
values of $\alpha ,$ let $n$ denote the integer part of $a=-\alpha
$. For this range of $\alpha $ we take
\begin{eqnarray}
({\cal D}_x^{a}f)(x)&=&{\frac 1{\Gamma (n-a +1)}}\\
 & &\times {\frac{d^{n+1}}{
dx^{n+1}}}\int_0^xdy\;(x-y)^{n-a}f(y).\nonumber
\end{eqnarray}
In this case 
\begin{equation}
{\cal D}_x^{a}x^\mu =\frac{\Gamma (\mu +1)}{\Gamma (\mu -a)}
t^{\mu -a },
\end{equation}
which is what one might expect for a derivative of order $a$.
Fractional derivatives can be related to anomalous diffusion by noting
that a diffusion process with $\langle x^2\rangle =Dt^\gamma $ can be
derived from a fractional Fokker--Planck equation through the use of
scaling arguments \cite{zas}.

There are many realizations of the fractional FP equations to physical
systems, including applications from turbulence to diffusion in
viscoelastic of porous media \cite{kzs}. Generally they are of the
form \cite{fp,fog,zas}
\begin{eqnarray}
{\cal D}^\delta_t  P(Q,t)&=& {\cal D}^\mu_{-Q}\left( A(Q)P(Q,t)\right) \\
& &+\frac 12{\cal D}^\nu_{-Q}\left(
B(Q)P(Q,t)\right) \nonumber
\end{eqnarray}
with fractional derivatives in space and/or time.  The powers
$\delta,\mu,\nu$ are chosen to reproduce anomalous diffusion through
scaling formulas such as $Q^2\sim t^{\gamma}$, where $\gamma$ is a
function of $\delta,\mu,\nu$.  But there are many limitations at the
moment to this type of phenomenological approach. First, one must
assume the existence of the coefficients $A$, $B$ and so forth, which
are now fractional moments. Further, since there is no microscopic
underpinning, extensions to higher dimensions become tenuous since
many more coefficients are needed, whose origin is then not
understood.

By starting with the quantum problem of a test particle in the chaotic
quantum environment, we have introduced certain `microscopic' scales
which describe the interactions between the systems. If we now take a
semi--classical limit of our influence functional, we will see that its
classical analog is a fractional Fokker--Planck equation. Further, the
quantum distribution functions are now solutions (in this limit) of
that equation.

The fractional Fokker--Planck equation is derived by taking the Wigner
transform $f(Q,P,t)$ of the density matrix $\rho (X,Y,t)$
\begin{eqnarray}
f(Q,P,t)&=&\int \frac {dR}{2\pi \hbar } \exp\left (
 -\frac{iPR}{\hbar}\right )\nonumber \\ 
 & & \times \rho \left( Q+\frac R2,Q-\frac R2,t\right) .
\end{eqnarray}
The classical transport equation is now defined for $f(Q,P,t)$
by Wigner transforming the evolution equation (19). To leading
order in $\hbar$,
$$
\frac{\partial f(Q,P,t)}{\partial t}=\int \frac{dR}{2i\pi
  \hbar ^2} \exp \left( -\frac{iPR} \hbar \right)
$$
\begin{eqnarray}
& &\left\{ -\frac{\hbar ^2}{2M}
\partial _Q\partial _R  +U\left( Q+\frac R2\right)-U\left(
  Q-\frac R2\right)\right. \nonumber\\
&& \left.  -i\Gamma^{\downarrow }\left| \frac R{X_0}\right|
  ^\alpha +i\gamma \hbar X_0\alpha {\rm sign}(R)
\left| \frac R{X_0}\right|^{\alpha -1}
\partial _R\right\}\nonumber\\
&&\times \rho \left( Q+\frac R2,Q-\frac R2,t\right) . 
\end{eqnarray}
We now use the Reisz fractional integro--differential operator
which is defined as \cite{samko,prl}
\begin{equation}
(-\Delta_P)^{\frac{\alpha }{2}} f(P) = {\cal F}^{-1} |X|^\alpha {\cal F} f(P),
\end{equation}
where $\Delta_P$ is the Laplacian (here with respect to the momentum
$P$), and ${\cal F}$ is a Fourier transform from $P$ to $X$.  For our
purposes we take $D^\alpha_P=(-i/\hbar)^\alpha
(-\Delta_P)^{\alpha/2}$, since $D^2_P[f]=\partial^2 f/\partial P^2$.
We also use $\overline{D}^\alpha_P=(-i/\hbar)^\alpha {\cal F}^{-1}
{\rm sign}(X)|X|^\alpha {\cal F}$, which has the property
$\overline{D}^1_P[Pf]=\partial (Pf)/\partial P$. The result is a
fractional Fokker--Planck equation in phase space, which is also a
fractional extension of Kramers equation:
$$
\frac{\partial f(Q,P,t)}{\partial t}  + \frac{P}{M}\frac{\partial f(Q,P,t)}{
\partial Q}-\frac{\partial U(Q)}{\partial Q}\frac{\partial f(Q,P,t)}{
\partial P}
$$
\begin{eqnarray}
   &&= \gamma _\alpha \left\{ \overline{D}^{\alpha
   -1}_P[Pf(Q,P,t)]\right.\\
   && \left.  -\frac{2TM}{\alpha\hbar^2}(i\hbar)^\alpha 
  D^\alpha_P[ f(Q,P,t)]\right\}.\nonumber
\end{eqnarray}
Here $T=1/\beta $ is the temperature and the operator and
the generalized friction coefficient is
\begin{equation}
\gamma_\alpha=\frac{\beta\Gamma^\downarrow\hbar\alpha}{2MX_0^\alpha}.
\end{equation}
The solutions to this equation follow directly from the Wigner
transform of our quantum solutions (20) and (24).

\section{Conclusions}

The quantum dynamics of test particles in complex quantum environments
was studied using influence functional methods. The stochastic
character of the dynamical evolution follows from the specific
properties of the chaotic/random matrix environment. Markovian
dynamics ranging from L\'evy to turbulent and to Brownian diffusion
are possible. In the classical limit, one can derive a fractional
Fokker--Planck equation, which reduces to the well--known Kramer's
equation for the special case of a complex environment. This provides
a connection between quantum chaos (RMT), stochastic processes and
fractional kinetic theory, which can be used to better understand the
origins of anomalous diffusion in complex classical and quantum
systems.


Table 1. Selected dynamical limits of the random matrix
influence functional of Ref. \cite{bdk2}.

\vspace{2mm}

\begin{tabular}{lll}
\hline\hline
 & & \\
 Classical Limit & $\alpha =2$ &  Kramer's Equation \cite{bdk2}  \\ 
 & & \\
                 & $\alpha <2$ &  Fractional Fokker--Planck
 Equation;  L\'evy Processes \cite{prl}\\
 & & \\
\hline
 & & \\
$U(x)=0$ & $\beta =0,\;\alpha =2$ &  Turbulent Diffusion \cite{pla}     \\ 
 & & \\
         & $\beta =0,\; \alpha <2$& Quantum L\'evy Processes \cite{prl}  \\ 
 & & \\
         &$\beta \neq 0$, BM limit         &     Quantum
 Brownian Motion with $\left\langle x^2\right\rangle =2Dt$ \cite{bdk2}\\
 & & \\
         &$\beta \neq 0,\; X_0\sim 1$  & $\left\langle
        x^{2}\right\rangle =2Dt$ but non--Maxwellian  distributions 
\cite{bdk2}.\\
 & & \\
\hline
 & & \\
 Weak Coupling & $\beta\ge 0$, $\alpha=2$  
  & Caldeira--Leggett Influence Functional \cite{bdk2,lw}\\
 & & \\
\hline\hline
\end{tabular}

\end{document}